# Towards Autonomous Cyber Operation Agents: Exploring the Red Case

**Li Li, Jean-Pierre S. El Rami, Ryan Kerr, Adrian Taylor, Grant Vandenberghe**
Defence Research & Development Canada
{li.li, jeanpierre.sabbaghelrami, ryan.kerr, adrian.taylor, grant.vandenberghe}@drdc-rddc.gc.ca

## Abstract

Recently, reinforcement and deep reinforcement learning (RL/DRL) have been applied to develop autonomous agents for cyber network operations (CyOps), where the agents are trained in a representative environment using RL and particularly DRL algorithms. The training environment must simulate CyOps with high fidelity, which the agent aims to learn and accomplish. A good simulator is hard to achieve due to the extreme complexity of the cyber environment. The trained agent must also be generalizable to network variations because operational cyber networks change constantly. The red agent case is taken to discuss these two issues in this work. We elaborate on their essential requirements and potential solution options, illustrated by some preliminary experimentations in a Cyber Gym for Intelligent Learning (CyGIL) testbed.

## 1 Introduction

Recently, deep reinforcement learning (DRL) has been applied to autonomous cyber operations (CyOps) [1-14]. The goal is to generate cyber defence blue agents that can reach a superhuman level of decision proficiency through automatic learning and which can defeat intelligent and experienced cyber attackers.

To achieve the strong blue agent, an intelligent red adversary is required. How may the technologically advanced foe, e.g., the AI-enabled red agent(s), attack our networks and systems? What should be the blue defensive countermeasure against such red attacks on the move? Indeed, the human blue and red team exercises are conducted routinely today to discover the potential attack vectors, to harden the networked systems, and to train the blue defensive operation capabilities employing crafted red attacks.

The autonomous red agent, sometimes called the autonomous red penetration-tester [1-5], presents one of the initial use cases for the research on applying DRL towards autonomous cyber operations (CyOps). The use case addresses the critical resource scalability issue for cyber red teams. A red team member needs years of training to be competent. The human red team exercises for network hardening and blue team training require weeks and months to prepare and execute, with outcomes dependent on the technical expertise of the red team member. The results may thus not provide measurable assurance on the network's defensive capabilities.

This work studies the autonomous red agent problem with the aim of network hardening and blue team training. The goal is a deployable red agent. The red problem is both simple and complex. The CyOps training environment for the red agent may be relatively simple to build because the limited red observation space contains only what the red can gather from the network. For example, it has the information discovered from the host the red agent resides on and scanned from the connected network nodes. However, due to the limited visibility and a large action space of various attack tactics, techniques, and procedures (TTP), the red problem for solving the optimized attack courses of action (CoA) is fundamentally difficult. An autonomous red agent is more valuable if DRL brings feasible and optimized solutions.

To achieve a deployable red agent, an effective agent training environment is first required. This is a good simulated DRL training environment, which represents network CyOps with high fidelity, and at the same time, enables efficient agent training to complete the sim-to-real loop for usable agents. As operational cyber networks change configurations constantly, generalizability of the trained agent is also critical for a feasible solution to prevent lengthy retraining [15].

With the increased interest in autonomous CyOps agents, a few training environments have been reported, open-sourced, and offered as public challenges [5-14]. To improve the realism of the simulated agent training environment, more detailed state-transition modelling of the cyber network has been employed [7,10,14]. The approaches require meticulous design and model abstraction to maintain implementation feasibility. The trained agents generally cannot operate the real CyOps tools in the network, nor handle the network topology changes [15]. In [12], a unified emulator and simulator training environment, namely the Cyber Gym for Intelligent Learning (CyGIL) was presented to generate a realistic simulated CyGIL-S automatically from the data collected from the emulated CyGIL-E. The solution was only reported with very limited test results [12-13]. At the same time, agent generalizability remains a critical problem.

We attempt to address these issues in this work, namely the CyOps agent training environment, and the network generalizability of the trained agent, focused on the red agent. The contributions are twofold. First, the presented CyGIL approach highlights some solution options in developing autonomous CyOps agents using DRL from sim-to-real. Second, the red agent generalizability studies demonstrate the potential leverage in red observation embedding for achieving the goal.

The rest of the paper is organized as follows. Section 2 presents the unified CyGIL emulator-simulator training environment and the representation learning for constructing CyGIL-S. The observation embedding scheme for supporting agent generalizability is discussed in Section 3. Section 4 presents some preliminary experimental results. Section 5 elaborates on the next steps and directions to conclude the paper.

## 2 CyOps Agent Training Environment

Network CyOps involve strategic decision-making for multi-stage operation actions towards an end goal. DRL provides a reasonable paradigm for this action decision problem. Modelled as the DRL agent, each red and blue agent decides its sequences of actions to maximize its final accumulated reward. The agent uses its observation of the network environment to choose actions. The algorithm iterates to refine the agent's policy model over many action steps, organized as episodes of the training game for the agent, which has the objective defined by the reward function and game-ending criteria.

### 2.1 System Model

To construct a CyOps agent training environment for any given cyber network, a generic framework is devised as depicted in Fig. 1. First, the cyber network and the blue and red CyOps tool sets form the environment and its blue and red action spaces, denoted in Fig.1 as action $a_B \in A_B$ and $a_R \in A_R$, where $A_B$ and $A_R$ are the blue and red agent action spaces, respectively. Then the data employed by the blue and red for conducting the operation substantiate the observation spaces, illustrated as $o_B \in O_B$, $o_R \in O_R$ where $O_B$ and $O_R$ are blue and red observation spaces, respectively. The data produced from the operation are used to calculate the rewards, denoted as $r_B$ and $r_R$ for the per step reward for the blue and red agents, respectively. The reward obtained at each step is generated by the reward functions that guide the CyOps objectives. The blue and red agents decide the action at each step by the trained policy $\pi(a_B|o_B)$ and $\pi(a_R|o_R)$ respectively.

CyGIL implements the framework by running the network and its CyOps tools on virtualized hardware to stand up CyGIL-E. In CyGIL-E, small networks are emulated on virtualized Mininet switches using the Open Network Operating System (ONOS) Software Defined Network (SDN) controller [8], and large networks using vSphere [17]. The CyOps action execution in CyGIL-E thus employs real blue and red team tools, e.g., the red team offensive tool CALDERA from MITRE [8,18].

As shown in Fig.1, the CyGIL library provides unified agent training across the emulator and simulator, to enable direct agent deployment in the real cyber network [12-13]. The simulated CyGIL-S is generated from the agent action trace data collected from CyGIL-E, including data used by and generated from the operational tools. CyGIL-E and CyGIL-S use the same data and thus can have the same representation of the action and observation spaces.

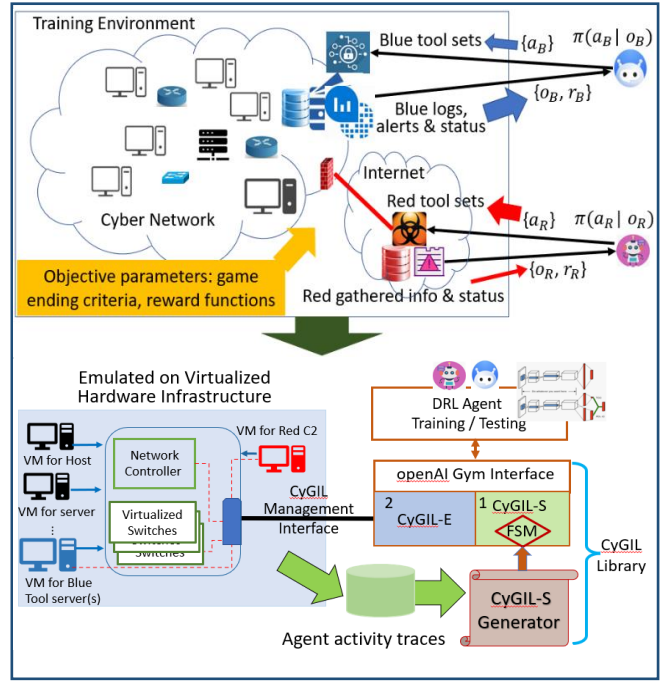

Fig. 1. CyOps training environment framework and CyGIL implementation: 1 – agent training in CyGIl-S; 2 – representation learning, agent transfer and verification in CyGIL-E; dashed lines – the interface between the CyGIL library and the network and CyOps tools; C2: Command and Control in the CyOps toolset(s)

The details of CyGIL-E and the unsupervised generation of CyGIL-S are presented in [8, 12]. As expected, CyGIL-S reduces the agent training latency from days and weeks to less than one hour [12]. CyGIL-S enables algorithm development, hyperparameter tuning, new objective training, and game parameter design, including, for example, reward function and game episode length tuning [12], all of which are not feasible in an emulator.

### 2.2 Representation Learning & Continual Agent Training

Data collection is very time-consuming in any real world network and physical systems [16]. It is crucial to collect as little data and as fast as possible to generate a sufficient CyGIL-S that embodies the state space distributions needed for the agent to reach the optimal policy attainable in the real network of CyGIL-E.

To this end, a unified CyGIL simulation environment generation and agent training algorithm is presented in [13]. The mechanism runs across CyGIL-E and CyGIL-S to conduct environment representation learning, continuous agent training and agent transfer, as illustrated in Fig. 2.

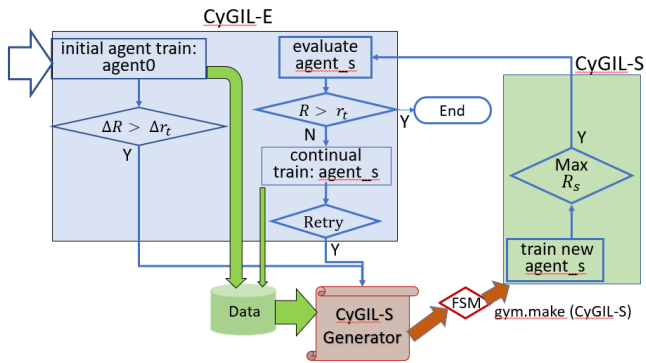

Fig. 2. Unified CyGIL-E and CyGIL-S

As in Fig. 2, for a given network scenario with its agent training objectives defined in a game, the training environment, i.e., the "gym" instance, is first stood up in CyGIL-E. An initial agent0 is trained in CyGIL-E to collect experience data traces. When $\Delta R$, the improvement of the accumulated rewards, referred to as the "return," jumps over a threshold $\Delta r_T$, the collected data are used to generate the FSM and the first CyGIL-S. At this stage, the training average return in CyGIL-E is often far below the optimal value, and the evaluation return may not yet show any increase. However, some good paths have already been traversed in the state space. The agent0 training in CyGIL-E uses the DRL algorithm for the representation learning of the state space. The DRL algorithm provides effective learning by moving towards better policy regions. Then the built CyGIL-S contains the state transitions relatively close to the solution area, even though covering only a very small part of the state space.

Next, the training game is opened in CyGIL-S to train new agent_s until its policy $R_s$ cannot be further improved. The trained agent_s is then transferred to CyGIL-E for evaluation. If the evaluation return $R$ is not satisfying, e.g., fluctuated returns below a predefined threshold value, the agent_s from CyGIL-S will be trained continuously for a few more shots in CyGIL-E to collect additional traces. This continuous training is a further environment representation learning taken by the agent_s by trying in the state space area closer to the optimal solution than the area where the data was collected last time in the CyGIL-E. This is because the agent_s model has already explored and learned better policy sets from previously collected data. The "Retry" condition can be as simple as a few more episodes, e.g., ten more episodes, as used in experiments presented in the next section. Then the collected traces are added to generate the next new CyGIL-S that trains a new agent_s. The process repeats until an agent_s trained in CyGIL-S produces satisfying evaluation results in the CyGIL-E; that is, the agent_s operates an optimized CoA in the emulated real network to achieve the defined CyOps objectives.

The agent_s currently is completely retrained in the new CyGIL-S each time because the latency is negligible compared with the emulator. This also allows for using different agent policy models for representation learning in different areas of the state space in CyGIL-E. For example, sample-efficient DRL algorithms at the beginning and more policy-accurate algorithms in the later stage. The "Retry" condition that stops the data collection in CyGIL-E to generate a new CyGIL-S is still empirically selected for different training scenarios and games. The final agent_s is intended to operate directly in the real (emulated) network with the optimal policy achieved from the training.

## 3 Generalization across Network Updates

As network updates occur constantly in operations, a usable agent must generalize to varied network configurations not seen during the training. The agent makes decisions based on its observation space (ObS). The data that form the observation space are collected by the agent from executing actions.

With modern red team tools, e.g., CALDERA® and Cobalt Strike®, the action execution and output data collection are managed by the tool's C2 (Command and Control). The red action is executed in the network by the implant(s), referred to as "hand(s)," which have landed on at least one of the network hosts through perimeter penetration, such as phishing and other initial access attacks. The hand is entirely controlled by C2, to which the hand beacons back to receive action execution commands and to report the result data collected from the execution. The collected data are used by C2 to parameterize and form the new action execution commands which are dispatched to the hands.

The collected data also form the ObS of the agent. Several observation space embeddings have been experimented with to improve agent generalizability across network changes, as illustrated in Fig. 3.

### 3.1 Full Information Embedding

The first CyGIL ObS in Fig. 3, the ACTNeT, consists of action parameters and the connectivity matrix, formatted as shown in Fig. 3(a), with the network size and action space limited by $N$ and $K$. The action parameters include those that are required for and acquired from all the several hundred CALDERA red actions. The host action parameters are associated with the host discovered. The net action parameters in the last Net Info Row (NIR) do not pertain to a particular host.

The current ACTNeT implementation has $K=25$, and $N =100$ to accommodate large networks of up to 100 hosts involved in the red attack. The 25 action parameters support CALDERA attack actions by recording the OS version, IP addresses, user and admin credentials, modifiable services, and certain types of processes, among others, about the hosts. The 101st row has network-wide parameters such as the domain name, domain admin credentials, etc.

In ACTNeT, one-hot encoding is applied to connectivity and many parameters, with the rest of the parameters encoded in integers to represent how many of them are collected. At the same time, all the detailed parameter values, e.g., IP addresses, are stored by CALDERA in its C2 database, as any typical state-of-the-art red tools do.

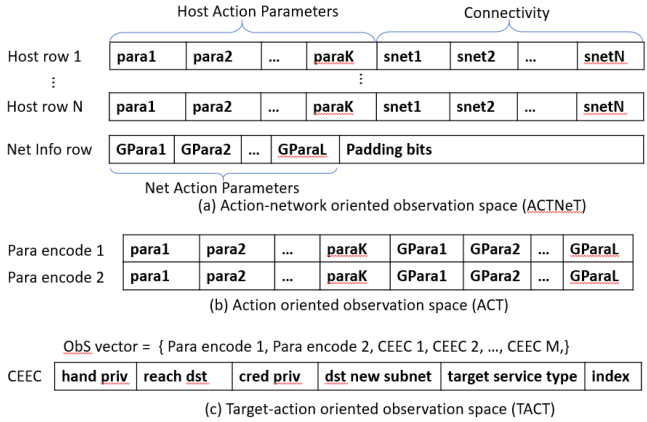

Fig. 3. ObS representations – (para: parameters; snet: subnet; GPara: network-wide parameters; priv: privilege; cred: credentials; dst: destination; reach: reachable; CEEC: command executable encoding)

ACTNeT is effective in training the red agent for different game objectives running on various networks, as demonstrated in [8, 12-13]. The agent policy also generalizes to network IP address variations, as address values are not in the ObS during training. However, the agent is not generalizable if new data values are brought about in the ObS due to network changes. This includes, for example, the host row order variation when the order in which the hosts are discovered in the network changes, the new subnet structure, and the number of instances collected for a parameter. The problem also arises when more or fewer hosts are found in the network, or more or fewer modifiable services are found on a host. In these cases, the trained agent fails because of the very different values in the ObS. However, in many of these cases, the trained policy still forms the optimal kill chain in the network when the same vulnerabilities are still presented and reachable, despite the network updates after the training. Irrelevant network details trap the agent. Training the agent through all the ObS variations to generalize is not preferable, given the largely increased training time and an unbounded number of alterations that may be encountered.

### 3.2 Action-Only Embedding

The action-oriented ObS (ACT) removes all ties to host and network specifics when embedding the action parameters, as shown in Fig. 3 (b). ACT only attends to action opportunities to provide network generalizability in the "smashing" kill-chain type of attack. The "para encode 1" indicates if the parameter is gathered, and the "para encode 2" flags if any new parameter instance has been collected by the action just executed, both using one-hot encoding.

The "smashing kill-chain" attack is to execute all the correctly formed action commands at all legit hands to spread, attack and impact as much as possible before being stopped. This is often the case in tactical scenarios and mission-oriented cyber networks. The red has little opportunity to gather enough parameters to form executable commands. The commands often fail even though parameterized correctly, caused by the low success rate, the peculiar conditions required and the largely unknown network conditions. This approach is robust in practice. First, the real opportunities for multiple hands and destinations are few, given the difficulties of red operation. Additionally, selecting one precise source and destination extends the training time enormously. The information collected about a host, as in the ACTNeT ObS, is inaccurate and may change. Therefore, the precise host selection reduces the already scanty red opportunities, often fails to reach an optimal solution in limited training time and is more susceptible to variations in real networks.

### 3.3 Action-Target Embedding

Can a red agent be trained to select a stealth action execution on a single preferred target host and still achieve network generalizability? To this end, the TACT embedding of the ObS, as shown in Fig. 3 (c), is being experimented with. TACT is a vector that extends the ACT by appending the command executable encodings (CEECs). Each CEEC encodes a parameterized, currently executable action command that is available for the agent to choose from. Currently, the "reach dst" and "dst new subnet" fields are one-hot encoded. The "target service type" marks if the command execution conditions are known to be met, in multiple bits, one for each condition. Some fields, for example, the destination reachability, as shown in "reach dst," and the condition matching level in the "target service type," can be inaccurate as they are often unknown and may change. Too many condition bits in the "target service type" reduce the generalizability. Each CEEC shows its action index in the action space. The trade-offs between more fields in the CEEC to benefit the precise selection and the achievable policy generalizability are still under investigation. The CEEC order and number invariant are examined in two options: the hand order managed by the red tool itself or the input position invariant properties of the attention modelling.

## 4 Experimental Results

The network, depicted in Fig. 4 with its variations, is applied in experiments. Hosts 1 and 2 in the network are reachable from the external "Internet" by the Attacker's C2. All hosts inside the network can reach the Active Directory (AD) Server/Domain Controller (DC) at Host 6, a Windows 2016 server. Other hosts run on Windows 10 (W) except Host 1 and Host 9 on Linux (L). Hosts on the same subnet switch can communicate with each other. Between different subnets, firewall rules allow Host 12 to communicate with Host 2 and Host 3 in addition to hosts in its subnet. Each host sends messages to at least one other host at any given time.

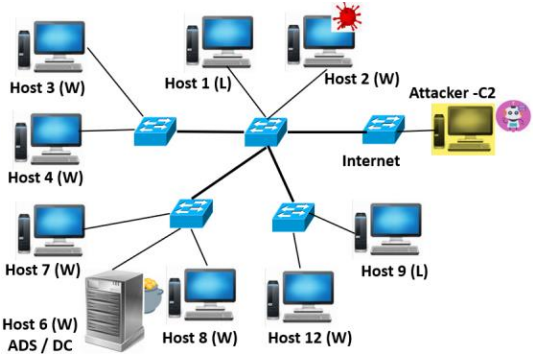

Fig. 4 Experimental Network

As the initial state of the training game, the red agent has already compromised Host 2 via phishing and implanted a "hand," i.e., malware. The hand reports back to C2 and receives action execution commands from C2. The red agent trained in CyGIL thus uses CALDERA and performs decision-making at C2 to organize the kill-chain sequence of actions and mobilize the hands to execute them, the same as today's human red team expert that uses CALDERA to conduct attacks. After the agent selects an action, CALDERA, by default, parameterizes and forms the action execution commands and sends them to all hands toward all destinations that are in the formulated commands.

The red agent action space is presented in Fig. 5, encompassing general tactics and techniques of several Advanced Persistent Threat (APT) groups [17] to enable the network end-to-end kill chain. The agent's objective is to land on the DC of ADS on Host 6. If it succeeds, the red agent will have the admin privilege to breach the entire domain.

The reward after each action is defined as $r = w - c$, where $c = 1$ is the cost and $w$ is the worth of the collected result. When the agent finally lands a hand on the DC of Host 6, $w = 100$, the episode ends. The game episode also ends when the agent executes 80 actions.

The network is emulated on a Windows machine running Intel(R) Core (TM) i9 with 64GB RAM. Agent training is performed on a DL laptop with a similar configuration. Running the CyGIL library for CyGIL-E and CyGIL-S, and the agent training algorithm, the DL laptop is connected to the emulated network using the VMware REST API over an Ethernet connection [8].

| ATT&CK Tactics | Actions - ATT&CK Techniques | Index |
|---|---|---|
| Discovery | T1135: Network Share Discovery | 0 |
| Discovery | T1087: Enumerate AD user accounts | 11 |
| Discovery | T1018: Remote System discovery | 12 |
| Discovery | T1016: Collect ARP details | 13 |
| Reconnaissance | T1590: Reverse lookup | 14 |
| Credential Access | T1003: Minikats to extract credentials | 7 |
| Credential Access | T1110: Brute force credentials | 10 |
| Privilege Escalation | T1548. Download & run Sandcat as admin | 8 |
| Lateral Movement | T1021: Sandcat remote fileshare WinRM | 1 |
| Lateral Movement | T1021: Sandcat remote fileshare (PsExec) | 2 |
| Lateral Movement | T1021: Sandcat with SCP (PsExec) | 3 |
| Lateral Movement | T1021: Sandcat remote using WinRM | 4 |
| Lateral Movement | T1021: Minikatz PSH and PsExec Sandcat remotely | 9 |
| Lateral Movement | T1021: Sandcat remote copy launch using PsExec | 15 |
| Lateral Movement | T1570: Tool transfer by WinRM and SCP | 6 |
| Lateral Movement | T1570: Tool transfer by file share | 5 |

Fig. 5 Red agent action space

## 4.1 Unified CyGIL-E and CyGIL-S

CyGIL-S enables the testing and selection of sample-efficient DRL algorithms. The C51 rainbow agent of Categorical Double Q Networks (CDQN) [21, 22] converges much faster than the DQN agent during the test in CyGIL-S, verified in evaluations in CyGIL-E, as illustrated in Fig.5.

The Proximal Policy Optimization (PPO) algorithm effectively converges to the optimal policy and has been applied in training the agent_s (Fig. 2) for representation learning and the final policy model. After comparing the CDQN and PPO in CyGIL-S and verified in CyGIL-E, CDQN is selected for its significantly faster convergence and equivalent optimal policy capability, as illustrated in Fig. 6.

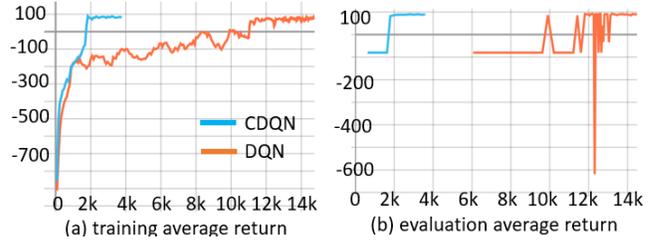

Fig. 5 Algorithm sample efficiency of C51 (CDQN) vs DQN; x-axis: training steps

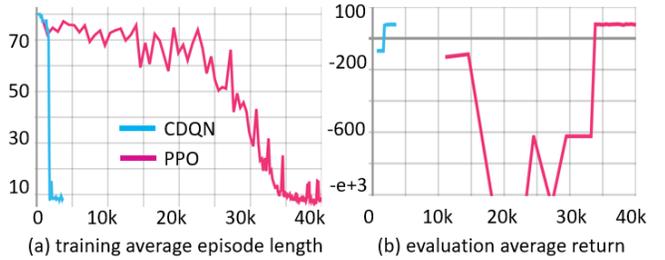

Fig. 6 Algorithm evaluation of C51 (CDQN) vs PPO; x-axis: training steps; episode length: the CoA with fewer actions to reach the goal is rewarded more in this training case.

The algorithm selection has improved the total agent training time. When trained entirely in CyGIL-E, as reported in [8], it takes from 1 week to 20 days for the DQN and PPO agent to reach optimal CoA in this scenario. The unified CyGIL-E and CyGIL-S reduce the time to a couple of days and even less than a day, as shown in Table 1.

| Full training in CyGIL-E | Continuous Training across E and S | |
|---|---|---|
| | DQN, PPO in E Final CDQN in S | CDQN, PPO in E Final CDQN in S |
| > 96 hours | ~ 38 hours | ~15 hours |

Table 1: Agent training latency in unified CyGIL-E and CyGIL-S

## 4.2 Agent Network Generalizability

The ACT ObS embedding is implemented in the CyGIL library and fully translatable to the original ACTNet. This enables the CyGIL-S to train agents using either ACT or ACT-

Net while CyGIL-E collects the full data set to support ACT-Net. The CyGIL-S can thus be built using ACTNet in the loop illustrated in Fig. 2. Then, the agent is trained in the final sufficient CyGIL-S on ACT. In preliminary tests, the agent training time increases by about 70% when using ACT instead of ACTNet in CyGIL-S. The training time in CyGIL-S is still under 1 hour. The PPO agent trained on the ACT is then sent to CyGIL-E to evaluate with the following network variations: (1) scramble IP addresses, including those of the hosts on the critical path; (2) scramble host names, IDs and order of their discovery, including the hosts on the critical path; (3) add and remove hosts while keeping the critical hosts; (4) move hosts on the critical path to different subnets. In all these network alterations, the vulnerabilities remain on the hosts even when they are moved in the network or have changed their host IDs or registered names in the network ADS and DC. Additionally, the connectivity between the hosts on the critical path remains in the firewall rules even though it may traverse through different subnets.

The agent trained in CyGIL-S in the original network of Fig.4 is sent to conduct 10 test runs in each of the four altered networks in CyGIL-E. The agent completes the attack following the optimal CoA in all the runs. The highest return the agent can receive is 92 if each action succeeds in one execution. The success ratios of actions are randomly distributed. The agent, therefore, often has to execute an action multiple times to make progress along the CoA.

The distributions of the evaluation returns are presented in Fig. 7, compared with 40 runs in the original trained network. The agent executes the trained policy and completes the CoA across all the varied networks. The return values lower than 92 are only caused by the extra steps when an action fails due to its statistical outcome. The lower return shown in the trained network is likely because the VMs are less stable after several hundred episodes of training and resetting. Then some actions need more repetitions to succeed.

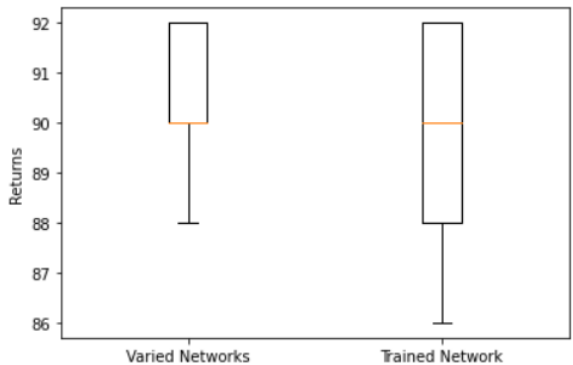

Fig. 7 Agent evaluation in CyGIL-E over network variations

## 5 Conclusions

This paper presents an approach that aims for a CyOps training environment that may develop applicable network CyOps agents. The work takes the red agent case because of the differences between the blue and red support. Two issues are studied, the sim-to-real agent training and transfer capability and agent generalizability when the operational network has changed after the training. The first set of preliminary test cases and results are presented.

CyGIL operates a unified training environment paradigm that enables the simulator auto-generation and continual agent training across the emulated CyOps network as CyGIL-E and its mirroring simulator of CyGIL-S. CyGIL-S allows the agent to be trained with high fidelity: agents trained in CyGIL-S use the same action and observation spaces that agents will encounter in real networks. The trained agent is, therefore, directly transferable to CyGIL-E. Training agents in CyGIL-S takes only a fraction of the time required in the real/emulated network. This allows for exploring various "what-if" scenarios that would otherwise be infeasible. The red observation space embedding is also studied to support agent generalizability when the operation network changes after the agent training. The ACT ObS embedding shows good generalizability for smashing kill-chain operations by trading off the precise target selection capability.

The initial results are promising for the tested scenario. However, the current approach is still experimental and requires further research to guide many detailed steps, such as the model transfer conditions between representation learning in CyGIL-E and agent policy training in CyGIL-S, as illustrated in Fig. 2: what parameters will trigger transfer between CyGIL-E and CyGIL-S and how many shots are needed in CyGIL-E; what training algorithms are more beneficial in different stages of the training cycle and how to select them without extensive tests, etc. The TACT ObS embedding for agent generalizability is still under investigation, where self-attention on currently parameterized action commands and CEEC invariability may provide the answer.

## References

[1] S. Chaudhary, A. O'Brien, and S. Xu, "Automated post-breach penetration testing through reinforcement learning", in Proceedings of 2020 IEEE Conference on Communications and Network Security (CNS), 2020.

[2] M. C. Ghanem and T. M. Chen, "Reinforcement Learning for Intelligent Penetration Testing", in Proceedings of Second World Conference on Smart Trends in Systems, Security and Sustainability (WorldS4), pp. 185–192. 2018

[3] H. Nguyen, H. N. Nguyen, and T. Uehara, "Multiple Level Action Embedding for Penetration Testing," The 4th International Conference on Future Networks and Distributed Systems (ICFNDS), 2020

[4] F. M. Zennaro and L. Erdodi, "Modeling penetration testing with reinforcement learning using capture-the-flag challenges and tabular Q-learning," arXiv preprint arXiv:2005.12632, 2020

[5] J. Schwartz and H. Kurniawati, "Autonomous Penetration Testing using Reinforcement Learning," CoRR, vol. abs/1905.05965, 2019.

[6] M. Sutana, A. Taylor and L. LI, "Autonomous network cyber offence strategy through deep reinforcement learning", in Proceedings of SPIE conference on Defences and Commercial Sensing, 2021, April 2021

[7] A. Molina-Markham, C. Miniter, B. Powell and A. Ridley, "Network Environment Design for Autonomous Cyberdefense," CoRR, vol. abs/2103.07583, 2021.

[8] L. Li, R. Fayad and A. Taylor, "CyGIL: A Cyber Gym for Training Autonomous Agents over Emulated Network Systems," in Proc. Autonomous Cyber Defence (ACD'21) workshop of International Joint Conference of AI , Montreal, Aug. 2021.

[9] Microsoft, CyberBattleSim Project - Documnent and source code, GitHub, 2021.

[10] M. Standen, M. Lucas, D. Bowman, T. J. Richer, J. Kim and D. Marriott, "CybORG: A Gym for the Development of Autonomous Cyber Agents," CoRR, vol. abs/2108.09118, 2021

[11] T. Kunz, C. Fisher, J, La Novara-Gsell, C. Nguyen, and L. Li, "A multiagent CyberBattleSim for RL cyber operation agents", in Proc. 9th International Conference on Computational Science and Computational Intelligence (CSCI), Las Vegas, Dec 2022

[12] L. Li, JP. S. El-Rami, J. H, Rao, A. Taylor, and T. Kunz, "Enabling a network AI gym for autonomous cyber agents", in Proc. 9th International Conference on Computational Science and Computational Intelligence (CSCI), Las Vegas, Dec 2022

[13] L. Li, JP. S. El-Rami, J. H, Rao, A. Taylor, and T. Kunz, "Unified emulation-simulation training environment for autonomous cyber agents", in Proc. 5th International Conference on Machine Learning for Networking, Paris, France, Nov 2022

[14] TTCP CAGE Challenges 2 & 3, on Github, 2023, https://github.com/cage-challenge

[15] J. Collyer, A. Andrew, and D. Hodges, "ACD-G: Enhancing autonomous cyber defense agent generalization through graph embedded network representation,", ML4Cyber workshop of the 39th International Conference on Machine Learning, Baltimore, Maryland, USA, Jul. 2022

[16] G. Dulac-Arnold, D. Mankowitz and T. Hester, "Challenges of Real-World Reinforcement Learning," 2019.

[17] MITRE Corporation, MITRE ATT&CK knowledge base, 2023.

[18] VMWare Vsphere product technical documentation https://docs.vmware.com/en/VMware-vSphere/index.html

[19] MITRE Corp., CALDERA - Document and source code, GitHub, 2023.

[20] OpenAI, Gym Documentation, https://www.gymlibrary.dev, 2022

[21] J. Farebrother, M. C. Machado and M. Bowling, "Generalization and Regularization in DQN," CoRR, vol. abs/1810.00123, 2018.

[22] M. G. Bellemare, W. Dabney and R. Munos, "A distributional perspective on reinforcement learning," in International Conference on Machine Learning, 2017.

[23] J. Schulman, F. Wolski, P. Dhariwal, A. Radford and O. Klimov, "Proximal Policy Optimization Algorithms," CoRR, vol. abs/1707.06347, 2017.